\documentclass[english,aps,pra,twocolumn,showpacs]{revtex4-1}
\usepackage[T1]{fontenc}
\usepackage[latin9]{inputenc}
\usepackage{xcolor}
\usepackage{amsmath}
\usepackage{amssymb}
\usepackage{graphicx}
\PassOptionsToPackage{normalem}{ulem}
\usepackage{ulem}
\usepackage{babel}

\begin{document}

\title{Franck-Condon Effect in Central Spin System}

\author{Li-Ping Yang,$^{1}$ Yong Li,$^{2}$}

\author{and C. P. Sun$^{1}$}

\email{suncp@itp.ac.cn}

\homepage{http://power.itp.ac.cn/~suncp}

\affiliation{$^{1}$State Key Laboratory of Theoretical Physics, Institute of
Theoretical Physics, Chinese Academy of Science, Beijing 100190, China\\$^{2}$Beijing
Computational Science Research Center, Beijing 100084, China}

\date{\today}
\begin{abstract}
We study the quantum transitions of a central spin surrounded by a
collective-spin environment. It is found that the influence of the
environmental spins on the absorption spectrum of the central spin
can be explained with the analog of the Franck-Condon (FC) effect
in conventional electron-phonon interaction system. Here, the collective
spins of the environment behave as the vibrational mode, which makes
the electron to be transitioned mainly with the so-called ``vertical
transitions'' in the conventional FC effect. The ``vertical transition''
for the central spin in the spin environment manifests as, the certain
collective spin states of the environment is favored, which corresponds
to the minimal change in the average of the total spin angular momentum. 
\end{abstract}


\maketitle

\section{INTRODUCTION}

The Franck-Condon (FC) principle, which determines the relative intensity
of the vibration-assisted electron transition spectrum, is of much
significance in molecular physics~\cite{Condon_1926,Herzberg}. In
these excitation and de-excitation processes, the transition probabilities
are proportional to the square of the overlap integrals between the
initial and final vibrational states (the FC factors). Compared with
the fast electronic transition, the vibrational motion is extraordinary
slow. As a result, during the electronic transition, vibrational coordinates
nearly keep stationary. This corresponds to a ``vertical transition''
picture on the effective vibrational potential energy surface, and
is called FC effect. The FC principle was originally proposed by Franck
to study the mechanism of photon-induced chemical reactions~\cite{Franck_1926}
and later expanded to the semi-classical formulation by Condon~\cite{Condon_1926}.
And then Lax applied this principle to solid-state physics~\cite{Lax_1952}.
However, all the previous works focused on electron-phonon coupling
system~\cite{FC_JPCS1954,FC_JCP1964,FC_PRB11,FC_NJP11_C. Lee,K. Huang_1952}. 

In this paper, we will study the FC effect induced by spin-spin interaction
for a model of a central spin in collective-spin environment, like
a central spin in quantum dot (QD)~\cite{QD_PRB_03,QD_PRL_02,QD_PRA1998,QD_PRB06,QD_Natrure05}
or in nitrogen-vacancy (NV) center~\cite{NV_PRL08,NV_PRL11_Zhaonan,NV_science06,NV_science08,NV_APL03,NV_APL10,NV_PRB01,NV_NJP08}.
In these systems, the unavoidable hyperfine interaction between the
central spin and the collective environmental spins is the chief culprit
of decoherence of the interested central spin. Thus, it is important
to investigate the effect of the spin-spin FC principle on the dynamics
of the central spin. On the other hand, in an ideal environment with
specific inter-spin coupling (such as Ising type), the central spin
can be used as a probe to explore the supersensitivity of a quantum
critical multi-spin system~\cite{PRA07_Zanardi,PRL06_Haitao Quan,PRA07_Haitao Quan}.
This theoretical prediction has been tested in several experiments~\cite{PRL08_Zhang,PRA09_J. F. Zhang,PRL10_J. F. Zhang}
and its robustness has been numerically shown as the longitudinal
field (equivalent to the tranverse hyperfine coupling in our model)
does not effect on the decoherence behaviour around the critical point~\cite{PRA09_J. F. Zhang,arxiv12_Ren Bao Liu}.

We consider the model of a central spin immersed in an environment
of nuclear spins. In general, the central spin can be a nuclear spin
or an electron spin. The central spin is initially polarized by the
crystal field in the $z$ direction. The collective environmental
spins behave as the vibrational mode in conventional electron-phonon
interaction model of FC effect. And the longitudinal hyperfine coupling
between the central spin and its spin environment is analogous to
the diagonal electron-phonon coupling, resulting in the effective
Hamiltonian of the environment spin being central-spin-dependent.
Due to this hyperfine coupling, when the central spin is excited by
the external field, the spin bath will be excited simultaneously and
this co-excitation generates the collective-spin-based FC effect.
And the FC factors, which were originally the overlap integrals between
the initial and final displaced vibrational Fock states, are defined
as the overlaps of the rotated collective spin states in our system.
An earlier paper investigated the spin FC effect, but it was only
devoted to demonstrating the Stokes shift in a spin-spin interaction
system~\cite{spin FC07}. In contrast to that work~\cite{spin FC07},
we study detailedly and systematically the collective-spin-based FC
effect and reveal its underlying physical mechanism.

We find that there exists the similar FC effect in our spin-spin interacting
system. In zero temperature case, the original Lorentz absorption
spectrum of a naked spin is shifted and split into few small peaks
by the weak hyperfine coupling, just as same as the vibronic transition
spectrum~\cite{electron_spectrum_JPCS1959}. The distribution of
the relative transition intensity is determined by the FC factors.
And the most probable transitions, which have largest FC factors,
are ruled by the ``vertical transition'' mechanism. On the other
hand, if the collective-spin environment is at finite temperature,
the peaks of the absorption spectrum of the central spin are depressed
and broadened significantly. Especially, when the hyperfine coupling
is strong enough, the excitation of the central spin is suppressed
intensively. This behavior is called FC blockade.

In the next section, we present our central spin model and its implementation
in N-V center in detail. In Sec.~III, we discuss the low excitation
limit of our central spin model and interpret the conventional FC
effect schematically. The collective-spin-based FC effect in our central
spin system are addressed in Sec.~IV. In Sec.~V, we study the collective-spin-based
FC effect with vertical transition in schematic perspective. Finally,
the summery of our main results is given in Sec.~VI. Some details
about the rotated Dicke state are displayed in Appendix.

\section{MODEL SETUP: CENTRAL SPIN IN ENVIRONMENT}

In the central spin system (e.g., QDs or NV center systems), the coupling
to the environmental nuclear spins destroys the coherence of the central
spin (electronic spin or large nuclear spin) primarily~\cite{central-spin_Nature11}.
While we find that these hyperfine couplings can arouse another interesting
effect\textemdash{}the collective-spin-based FC effect.

We consider a general model of a central spin immersed in an $N$-spin
environment, with the Hamiltonian described by $(\hbar=1)$ 
\begin{equation}
H=\mathcal{H}_{s}\left(S_{z}\right)+A\sum_{j=1}^{N}S_{z}\cdot I_{z}^{\left(j\right)}+\omega_{\mathrm{nu}}\sum_{j=1}^{N}I_{x}^{\left(j\right)}+\omega_{\mathrm{el}}S_{x},\label{H}
\end{equation}
where $S_{\alpha}$ and $I_{\alpha}^{\left(j\right)}$ ($\alpha=x,y,z$)
are the angular momentum operators of the central spin and the $j$th
nuclear spin, respectively. By virtue of the existence of the crystal
field, the central spin is ususally polarized in a particular direction
($z$ direction). As a result, $\mathcal{H}_{s}\left(S_{z}\right)$
generally depends on $S_{z}$, e.g., for NV center system, $\mathcal{H}_{s}\left(S_{z}\right)=DS_{z}^{2}$.
When a transverse external field along the $x$ axis is applied, the
central and environmental spins get additional Zeeman splittings $\omega_{\mathrm{el}}$
and $\omega_{\mathrm{nu}}$, respectively. As we know, the last term
in Eq.~(\ref{H}), $\omega_{\mathrm{el}}S_{x}$ will induce transitions
between the ground and excited states of the central spin. Because
of the existence of the spin environment, these transitions will be
modulated by the longitudinal hyperfine coupling $A(>0)$. For simplicity,
the hyperfine coupling between the central spin and the environment-spin
ensemble has been assumed to be homogeneous. 

It is convenient to define the collective polarization operators for
the nuclear spin ensemble as
\begin{equation}
J_{\alpha}=\sum_{j}I_{\alpha}^{\left(j\right)}.
\end{equation}
It is ready to find that these collective operators satisfy the following
commutation relations:
\begin{equation}
[J_{\alpha},J_{\beta}]=i\epsilon_{\alpha\beta\gamma}J_{\gamma},\ [J^{2},J_{\alpha}]=0,
\end{equation}
with $\epsilon_{\alpha\beta\gamma}$ the totally antisymmetric Levi-Civita
tensor. As a result, the collective environmental spins have collective
eigenstates (i.e., Dicke state~\cite{Dicke state-1,Dicke states-2})
and the dynamic symmetry is described by the algebra SO~$(3)$.

\begin{figure}
\includegraphics[width=8cm]{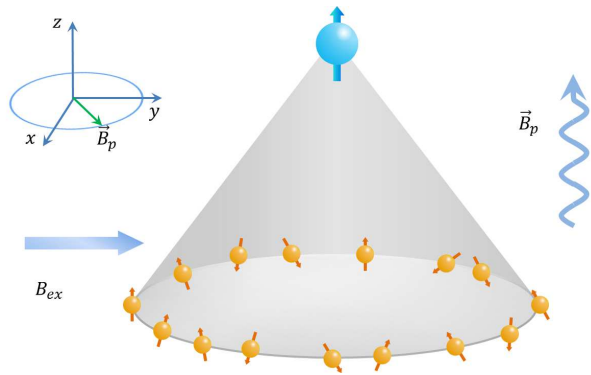}

\caption{(Color online) \label{fig:Schematic}Schematic of central spin model.
The central spin (the top blue one) is polarized in the $z$ direction.
The flipping of the central spin induced by the polarized field $\vec{B}_{p}$
is modulated by the longitudinal hyperfine coupling between the central
spin and its nuclear-spin environment.}
\end{figure}

Now let us focus on the model of a central electronic spin implemented
in the negatively charged NV center in diamond (Fig.~\ref{fig:Schematic}).
The electronic ground state of the center, which we concern, is a
spin triplet ($S=1$) and there is zero-field splitting (ZFS) $D=2.87\ \mbox{GHz}$
between states $\left|s=0\right\rangle $ and $\left|s=\pm1\right\rangle $
($\left|s\right\rangle $ is the eigenstate of $S_{z}$). It should
be noted that the $\left|s=\pm1\right\rangle $ levels are degenerate.
Neglecting the transverse hyperfine coupling between the electronic
spin and its environmental nuclear spin ensemble, we rewrite our model
Hamiltonian as $H=H_{0}+H_{1}$, with
\begin{equation}
H_{0}=DS_{z}^{2}+\omega_{\mathrm{nu}}J_{x}+AS_{z}J_{z},\label{H0}
\end{equation}
and
\[
H_{1}=\omega_{\mathrm{el}}S_{x}+\Omega\left(S_{x}\cos\omega t+S_{y}\sin\omega t\right).
\]
The applied static magnetic field is weak $B_{\mathrm{ex}}=12\ \mathrm{Gauss}$
and the corresponding Zeeman splittings of the electronic and nuclear
spins are $\omega_{\mathrm{el}}=g_{e}\mu_{B}B_{\mathrm{ex}}\approx211.35\ \mathrm{MHz}$
and $\omega_{\mathrm{nu}}=g_{n}\mu_{n}B_{\mathrm{ex}}\approx0.15\mathrm{MHz}$,
respectively~\cite{NV_science08}. We have taken the isotropic Land$\acute{\mbox{e}}$
$g$-factor of the electron and nuclear as $g_{e}\approx2.0$ and
$g_{n}=1.4$, respectively. $\mu_{B}(\mu_{n})$ is the Bohr (nuclear)
magneton and $\mu_{n}\approx10^{-3}\mu_{B}$. The hyperfine coupling
constant $A$ is of the order of $(\mathrm{kHz}\textrm{--}\mathrm{MHz})$.
And in order to probe the absorption spectrum of the central spin,
a circularly polarized field $\vec{B}_{p}=(B_{p}\cos\omega t,B_{p}\sin\omega t,0)$
is added with $\Omega=g_{e}\mu_{e}B_{p}=D/20$. Since $\omega_{\mathrm{el}},\Omega\ll D$,
we take these terms as perturbations. The frequency $\omega$ of the
circularly polarized field is required to be nearly resonant with
the ZFS of the central electronic spin, i.e., $\omega\approx D$.
Under the rotating wave approximation, the $H_{1}$ part reduces to
\begin{equation}
H_{1}\approx\frac{\Omega}{\sqrt{2}}\left[\left|1\right\rangle \left\langle 0\right|e^{-i\omega t}+\left|0\right\rangle \left\langle 1\right|e^{i\omega t}\right],\label{H1}
\end{equation}
where we have used the condition $\omega_{\mathrm{el}},\Omega\ll D$
as well as $\omega\approx D$.

\section{LOW EXCITATION APPROXIMATION }

\begin{figure}
\includegraphics[width=6cm]{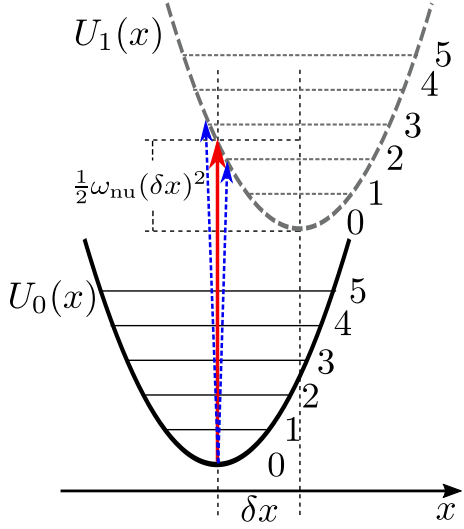}

\caption{\label{fig:vertical transition-1}(Color online) Schematic of the
conventional vertical transition. }
\end{figure}

To see why the above system enjoys the FC effect, we first consider
the semi-classical interpretation for the conventional FC principle
in an electron-phonon interacting system: in the approximation of
the linearization of the vibronic coupling, the phonons will get electron-dependent
displaced effective potentials (see Fig.~\ref{fig:vertical transition-1}).
The electronic transition is so fast that the geometry structure of
the vibrational freedom of degrees remain unchanged during this process,
i.e., vertical transition takes place most possibly. This conventional
FC phenomenon could be described by a simple model of a two-level
system coupling to a single vibrational mode, with the Hamiltonian,

\begin{equation}
H=\frac{\varepsilon}{2}\sigma_{z}+\omega_{p}b^{\dagger}b+g\sigma_{z}(b^{\dagger}+b),
\end{equation}
where $\sigma_{z}$ is the Pauli operator, $\varepsilon$ is the energy
difference between the ground and excited states of the eletron, $b^{\dagger}(b)$
is the bosonic creation (annihilation) operator of the vibration mode
with frequency $\omega_{p}$, and $g$ is the elctron-phonon coupling
constant.

Next we will show that our central electronic spin model can be reduced
into the above model in low excitation limit. In the subspace spanned
by the states $\left\{ \left|N/2,m\right\rangle ,\ m=0,1,2\dots,N\right\} $,
where $\left|N/2,m\right\rangle $ is the eigenstate of $\{J^{2},\ J_{x}\}$,
one can use the Holstein-Primakoff (HP) transformation
\begin{equation}
\begin{cases}
\tilde{J}_{+}\equiv J_{z}-iJ_{y}=b^{\dagger}\sqrt{N-b^{\dagger}b},\\
\tilde{J}_{-}\equiv J_{z}+iJ_{y}=\sqrt{N-b^{\dagger}b}b,\\
J_{x}=b^{\dagger}b-\frac{N}{2},
\end{cases}\label{H-P}
\end{equation}
with defining the bosonic creation and annihilation operators $b^{\dagger}$
and $b$. In the low excitation limit $\langle b^{\dagger}b\rangle\ll N$,
HP transformation (\ref{H-P}) can be given, expanded to the lowest
order, as
\begin{equation}
\tilde{J}_{+}\approx\sqrt{N}b^{\dagger},\ \tilde{J}_{-}\approx\sqrt{N}b,\ J_{x}=b^{\dagger}b-\frac{N}{2}.
\end{equation}

Then we obtain the Hamiltonian $\tilde{H}=\tilde{H}_{0}+\tilde{H}_{1}$
of a electron-phonon-like interaction model with
\begin{equation}
\tilde{H}_{0}=DS_{z}^{2}+\omega_{\mathrm{nu}}b^{\dagger}b+\frac{1}{2}\sqrt{N}AS_{z}\left(b^{\dagger}+b\right)-\frac{N}{2}\omega_{\mathrm{nu}},\label{H0-1}
\end{equation}
and the perturbation part $\tilde{H}_{1}=H_{1}$ describing the transition
between the electronic spin states $\left|s=0\right\rangle $ and
$\left|s=1\right\rangle $. The $\tilde{H}_{0}$ part can be diagonalized
by the displaced Fock state $\left|\xi_{s},m;s\right\rangle =\exp\left(\xi_{s}b^{\dagger}-\xi_{s}^{*}b\right)\left|m\right\rangle \otimes\left|s\right\rangle $~\cite{displaced fock state1,displaced fock state2},
where $s$ denotes the eigenstate of $S_{z}$, $\left|m\right\rangle \ (m=0,1,...,N)$
is the $m$-th Fock state of $b^{\dagger}b$, and the electron-dependent
displacement
\begin{equation}
\xi_{s}=-\xi_{s}^{*}=-\sqrt{N}\frac{sA}{2\omega_{\mathrm{nu}}}.
\end{equation}
The FC factor of the transition from $\left|\xi_{0},m;0\right\rangle $
to $\left|\xi_{1},n;1\right\rangle $ is defined as the overlap integral
between the two relative displaced Fock states 
\begin{eqnarray}
\left\langle \xi_{1},n\mid\xi_{0},m\right\rangle  & \!=\! & e^{-\left|\xi\right|^{2}/2}\!\!\sqrt{\frac{m!}{n!}}L_{m}^{n-m}\left(\left|\xi\right|^{2}\right)\left(\xi\right)^{n-m}\!\!\!\!,\label{F-C factor}
\end{eqnarray}
with $\xi=\xi_{0}-\xi_{1}$ and $L_{m}^{n-m}\left(x\right)$ being
the generalized Laguerre polynomial. 

We can formally define dimensionless canonical coordinates of the
vibrational mode as $x\equiv\left(b^{\dagger}+b\right)/\sqrt{2}$
and $p\equiv i\left(b^{\dagger}-b\right)/\sqrt{2}$, with $[p,x]=-i$.
The effective potentials of the nuclear spins are different correponding
to different eigenstates ($\left|s\right\rangle $) of the central
spin: 
\begin{equation}
U_{s}\left(x\right)=\frac{1}{2}\omega_{\mathrm{nu}}x^{2}+\sqrt{\frac{N}{2}}sAx-\frac{N}{2}\omega_{\mathrm{nu}}.
\end{equation}
As we know, if the electron spin and the nuclear spins are both in
the ground state $\left|\xi_{0},0;0\right\rangle $ initially, when
the electron spin is exited by $\tilde{H}_{1}$ to its higher level
$\left|1\right\rangle $, the nuclear spins could be co-excited to
all possible eigenstates $\left|\xi_{1},n;1\right\rangle $ of $U_{1}\left(x\right)$.
But the transition, which has the largest FC factor, is most favored.
From Eq.~(\ref{F-C factor}), we find that the most favored final
state $\left|n_{\mathrm{mf}}\right\rangle $ is determined by
\begin{equation}
n_{\mathrm{mf}}=\left[\frac{1}{2}\left(\delta x\right)^{2}\right].
\end{equation}
Here $\delta x=\sqrt{2}\xi$ and $\left[\cdot\cdot\cdot\right]$ means
the rounding operation. As shown in Fig.~\ref{fig:vertical transition-1},
the most favored transition coincides with the vertical transition:
the sole transition (red solid arrow) or two adjacent transitions
(blue dashed arrows). As a consequence, when the system is initially
in the ground state, vertical transition takes place most possibly. 

Thus, in the present spin-spin interaction system, there exists conventional
FC effect in low excitation limit.

\section{COLLETIVE-SPIN-BASED FRANCK-CONDON EFFECT}

In the previous section, we have shown that our central spin model
is approximately equivalent to an electron-phonon interaction model
in low excitation limit. In this section, we investigate the influence
of the nuclear spin ensemble on the transition spectrum of the central
spin and find collective-spin-based FC effect exists in this central
spin system.

The Hamiltonian $H_{0}$ is diagonalized in its direct product Hilbert
space as: $H_{0}^{\left(s\right)}\otimes\left|s\right\rangle \left\langle s\right|$.
Here $H_{0}^{\left(s\right)}$ is the effective Hamiltonian of the
nuclear-spin environment with the central spin at state $\left|s\right\rangle $
and given by 
\begin{equation}
H_{0}^{\left(s\right)}\equiv sAJ_{z}+\omega_{\mathrm{nu}}J_{x}+s^{2}D.\label{H0_eff}
\end{equation}
The eigenstate of $H_{0}^{\left(s\right)}$ is the rotated Dicke state
$\left|\theta_{s},m\right\rangle =\exp\left(-i\theta_{s}J_{y}\right)\left|m\right\rangle $
(please refer to Appendix A), where $\left|m\right\rangle $ is the
eigenstate of $J_{z}$ with eigenvalue $(m-N/2)$ and the mixing angle
is determined by 
\begin{eqnarray}
\sin\theta_{s} & = & \frac{\omega_{\mathrm{nu}}}{\tilde{\omega}_{\mathrm{nu}}(s)},\ \cos\theta_{s}=\frac{sA}{\tilde{\omega}_{\mathrm{nu}}(s)},
\end{eqnarray}
with
\begin{equation}
\tilde{\omega}_{\mathrm{nu}}(s)=\sqrt{\omega_{\mathrm{nu}}^{2}+\left(sA\right)^{2}}.
\end{equation}
As a result, we obtain the total eigenstate of $H_{0}$ as $\left|\theta_{s},m;s\right\rangle =\left|\theta_{s},m\right\rangle \otimes\left|s\right\rangle $
and the corresponding eigenvalue $E_{s,m}=s^{2}D+\left(m-N/2\right)\tilde{\omega}_{\mathrm{nu}}\left(s\right)$
for $s=0,\pm1$ and $m=0,1,2,...,N$.

\subsection{Franck-Condon factor}

As we know, the conventional FC effect describes the phenomenon that,
in the electron-phonon interacting system, the electronic transition
is modulated by the vibronic coupling. The transition probability
is proportional to the square of the FC factor, which is specifically
defined as the overlap integral between the two vibrational states
involved in the transition. Here we can also obtain the counterpart
of such factor in spin-spin coupling system as the overlap of two
rotated Dicke states:
\begin{eqnarray}
f_{m\rightarrow n} & \equiv & \left\langle \theta_{1},n\mid\theta_{0},m\right\rangle =d_{n,m}^{N/2}\left(\theta\right),\label{F-C_factor1}
\end{eqnarray}
with the rotation-angle difference $\theta_{0}-\theta_{1}\equiv\theta\in(0,\pi/2)$
and the element of the Wigner's (small) d-matrix
\begin{eqnarray}
d_{n,m}^{N/2} & = & \left[m!(N-m)!n!\left(N-n\right)!\right]^{\frac{1}{2}}\nonumber \\
\times\!\!\!\! &  & \!\!\!\!\sum_{k}\!\!\frac{\left(-\right)^{k}\!\!\left(\!\cos\frac{\theta}{2}\!\right)^{N+m-n-2k}\!\!\left(\!-\sin\frac{\theta}{2}\!\right)^{n-m+2k}}{\left(N-n-k\right)!\left(m-k\right)!\left(k+n-m\right)!k!}.\label{d-matrix}
\end{eqnarray}
For the special transition from $\left|\theta_{0},0;0\right\rangle $
to $\left|\theta_{1},n;1\right\rangle $, the FC factor reduces to
a simple form 
\begin{eqnarray}
f_{0\rightarrow n} & = & \sqrt{\frac{N!}{\left(N-n\right)!n!}}\left(\cos\frac{\theta}{2}\right)^{N-n}\left(-\sin\frac{\theta}{2}\right)^{n}.\label{F-C_factor2}
\end{eqnarray}

\begin{figure}
\includegraphics[width=7cm]{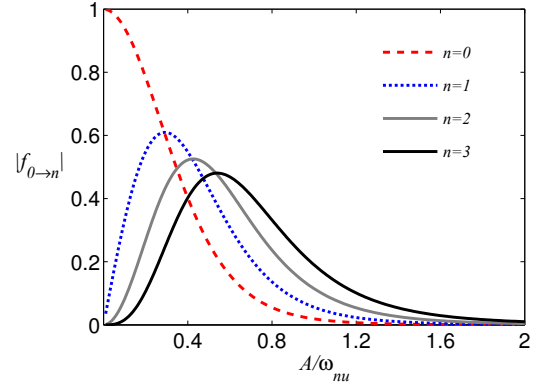}

\caption{\label{fig:F-C factor(m=00003D0)}(Color online) The amplitude of
the Franck-Condon factor vs the hyperfine coupling $A$ for transition
from $\left|\theta_{0},0;0\right\rangle $ to $\left|\theta_{1},n;1\right\rangle $.}
\end{figure}

As depicted in Fig.~\ref{fig:F-C factor(m=00003D0)}, for a specific
final state $\left|\theta_{1},n;1\right\rangle $, the amplitude of
the FC factor $f_{0\rightarrow n}$ has its sole maximum and the corresponding
hyperfine coupling strength to the maximum, $A_{n}^{\mathrm{max}}$,
increases gradually with $n$. But the maximum of $|f_{0\rightarrow n}|$
decreases with $n$ and the FC factor is drastically suppressed in
the case of strong coupling $A$. It is worth noting that, for a given
coupling strength $A$, there exists a sole or two adjacent most favored
collective-spin states$\left|\theta_{1},n\right\rangle $, which have
largest FC factors. This is one of the essences lying at the core
of the FC principle, since it will correspond to a very intuitive
physical picture\textemdash{}vertical transition.

\begin{figure}
\includegraphics[width=9cm]{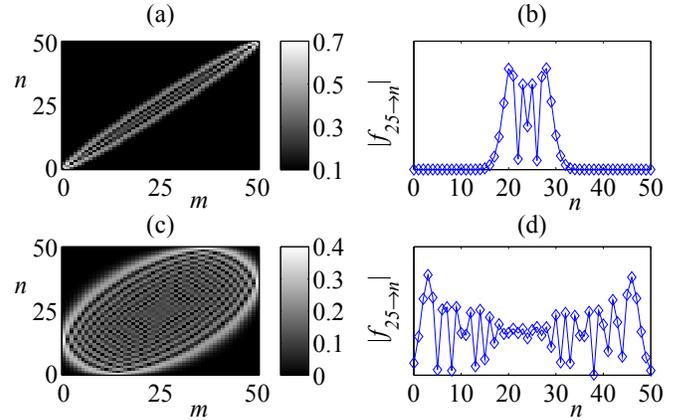}\caption{\label{fig:F-C_factorS}(Color online) Franck-Condon factor for transition
from $\left|\theta_{0},m;0\right\rangle $ to $\left|\theta_{1},n;1\right\rangle $
with different hyperfine couplings. (a) $A=0.2\omega_{\mathrm{nu}}$.
(b) $A=0.2\omega_{\mathrm{nu}}$ and $m=25$. (c) $A=2\omega_{\mathrm{nu}}$.
(d) $A=2\omega_{\mathrm{nu}}$ and $m=25$.}
\end{figure}

Different from $f_{0\rightarrow n}$, the monotonicity of $f_{m\rightarrow n}$
($m\neq0$) is much complicated, sine there are multiple local maximums
for a given $A$. We numerically calculate the FC factors of the all
the possible transitions for coupling constant $A=0.2\omega_{\mathrm{nu}}$
or $A=2\omega_{\mathrm{nu}}$, as shown in Fig.~\ref{fig:F-C_factorS}.
It is found that neighboring transitions ($\left|m-n\right|$ is small)
have larger FC factors when the hyperfine coupling is weak, but these
transitions are depressed in strong coupling case.

\subsection{Transition probability and absorption spectrum}

Starting from an initial state $\left|\theta_{0},m;0\right\rangle $,
the probability of the system in the state $\left|\theta_{1},n;1\right\rangle $
at time $t$ is given by
\begin{eqnarray}
P_{1,n;0,m}\left(t\right) & = & \frac{\Omega{}^{2}}{2}\times F(\frac{\omega_{1n,0m}-\omega}{2},t)\left|f_{m\rightarrow n}\right|^{2},\label{transition_probability}
\end{eqnarray}
where we have defined $\omega_{s'n,sm}=E_{s',n}-E_{s,m}$ and $F(\omega,t)=\left(\sin^{2}\omega t\right)/\omega^{2}$.
With the Fermi's golden rule, the transition rate of the central spin
from $\left|\theta_{0},m;0\right\rangle $ to $\left|\theta_{1},n;1\right\rangle $
is obtained as 
\begin{equation}
k_{m\rightarrow n}\left(\omega\right)=\lim_{t\rightarrow\infty}\frac{P_{1,n;0,m}\left(t\right)}{t}.\label{transition_rate}
\end{equation}

\begin{figure}
\includegraphics[width=7cm]{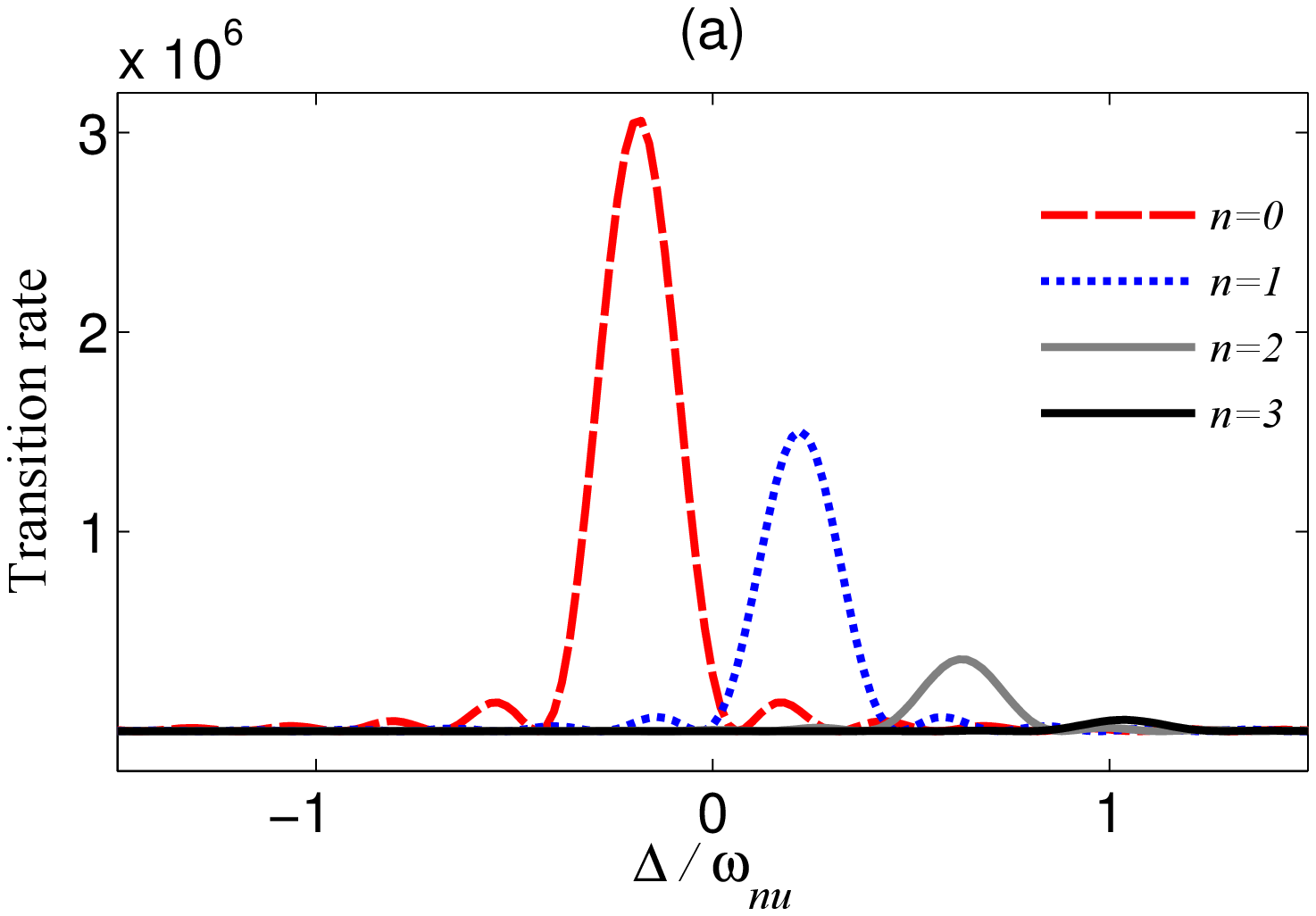}

\includegraphics[width=7cm]{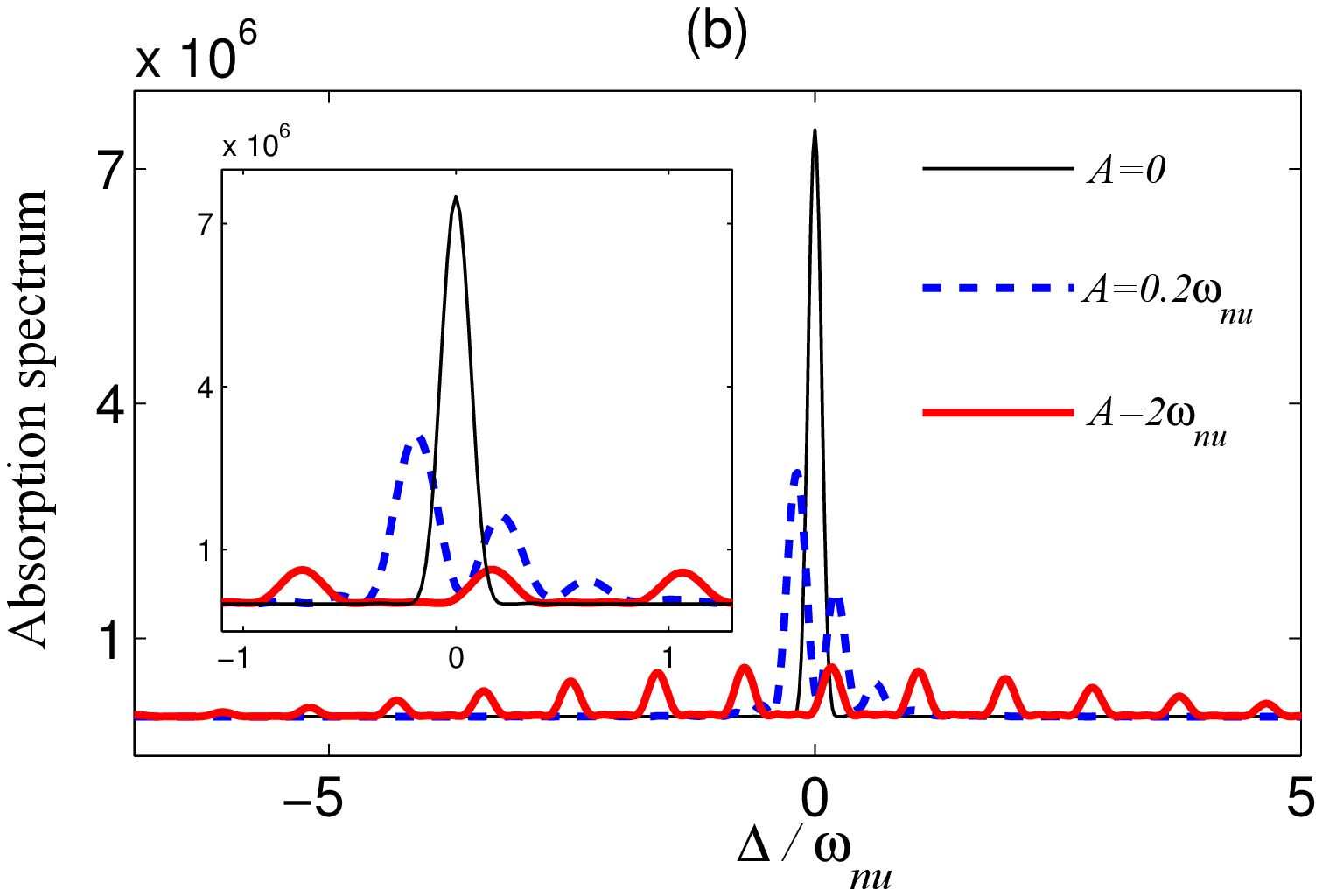}

\caption{\label{fig:ransition-rates and spectrum}(Color online) (a) Transition
rates $k_{0\rightarrow n}$ for system from the state $\left|\theta_{0},0;0\right\rangle $
to different final states $\left|\theta_{1},n;0\right\rangle $ vs
$\Delta$ ($\Delta=\omega-D$), where the hyperfine coupling $A=0.2\omega_{\mathrm{nu}}$
and the number of the nuclear spins $N=50$. (b) The absorption spectra
of the central spin with different hyperfine couplings.}
\end{figure}

Now we first consider the zero temperature case. The system is initially
in the ground state $\left|\theta_{0},0;0\right\rangle $. It is found
that the transition rates vary with the different final states. As
shown in Fig.~\ref{fig:ransition-rates and spectrum}~(a), for the
weak coupling $A=0.2\omega_{\mathrm{nu}}$, there exist a favored
transition, i.e., the one from the ground state to $\left|\theta_{1},0;1\right\rangle $
. From Eqs.~(\ref{transition_probability}) and ~(\ref{transition_rate}),
we know that the transition rate is not only determined by the Fermi's
golden rule, but also modified by the FC factor $f_{0\rightarrow n}$.

\begin{figure}
\includegraphics[width=7cm]{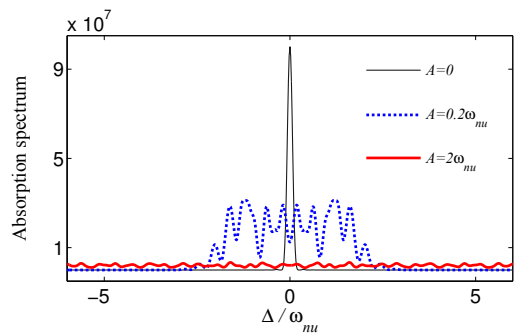}

\caption{\label{fig:Spectrum_T=00003D300K}Spectrum of the central spin at
room temperature with different hyperfine couplings $A$.}
\end{figure}

By summing up the transition rates over all the final states, we obtain
the absorption spectrum as 
\begin{equation}
I_{s}(\omega)=\sum_{n}k_{0\rightarrow n}\left(\omega\right).
\end{equation}
It should be noted that when the time $t\rightarrow\infty$, we have
this limit
\begin{equation}
\lim_{t\rightarrow\infty}\frac{F(\omega,t)}{t}=2\pi\delta(\omega).
\end{equation}
Then the spectrum becames to many width-less lines. Actually, the
observed absorption lines from experiments must have finite width
resulting from the interaction of the radiation field on the central
spin or any other interactions. However, the line-width correction
is neglected, since the positions and the relative heights are mostly
concerned and the linewidths of all the absorption lines are the same
in our case. And we take a finite time $\omega_{\mathrm{nu}}t=10$
in Eq.~(\ref{transition_rate}) in our numerical calculations in
Figs.~\ref{fig:ransition-rates and spectrum} and~\ref{fig:Spectrum_T=00003D300K}.
As we know, the absorption spectrum of a naked electronic spin ($A=0$)
given by the Fermi's golden rule is of the Lorentz form peaked at
$\Delta\equiv\omega-D=0$ (the thin black line in Fig.~\ref{fig:ransition-rates and spectrum}~(b)).
The Lorentz spectrum is shifted and split into small peaks, as a result
of the coupling between the electron spin and its nuclear-spin environment.
When the coupling is weak $A=0.2\omega_{\mathrm{nu}}$, the spectrum
is just split into few peaks neighboring the original one. However,
in strong coupling case, the absorption spectrum is composed by many
small discrete peaks centered at $\Delta=0$. Hence, the excitation
of the central spin introduced by a polarization field is inhibited
by the strong hyperfine coupling. And this phenomenon of transition
suppression was called ``Franck-Condon blockade''~\cite{FC_blockade_PRL05,FC_blockade_PRB06}.

In the case of finite ambient temperature, the collective nuclear
spins are usually in the thermal equilibrium state
\begin{equation}
\rho_{\mathrm{nu}}=\frac{1}{Z}\sum_{m=0}^{N}\exp\left(-\frac{m\omega_{\mathrm{nu}}}{k_{B}T}\right)\left|\theta_{0},m\right\rangle \left\langle \theta_{0},m\right|,
\end{equation}
where $Z=\sum_{m=0}^{N}\exp\left(-m\omega_{\mathrm{nu}}/k_{B}T\right)$
is the partition function, $k_{B}$ is the Boltzmann constant, and
$T$ is the temperature of the environment. Since the energy splitting
of the nuclear spins due to the weak external static field is small
(i.e., $\omega_{\mathrm{nu}}\ll k_{B}T$), all the Dicke state $\left|\theta_{0},m\right\rangle $
are nearly equally occupied in the case of high temperature. 

The absorption spectrum of the central spin at room temperature $T=300\ \mathrm{K}$
is shown in Fig.~\ref{fig:Spectrum_T=00003D300K}. For small $A$,
only the neighboring transitions, which possess large FC factors,
contribute significantly to the spectrum. Incorporated the Fermi's
golden rule (i.e., the energy-conservation requirement $\omega_{1n,0m}-\omega=0$),
the absorption peaks are close to the resonant point $\Delta=0$.
Nevertheless, when the hyperfine coupling is large, the relative intensity
of all the possible transition are almost the same. Consequently,
the excitation of the central spin is drastically suppressed due to
the Franck-Condon blockade.

\section{SCHEMATIC PERSPECTIVE VIEWS FOR FRANCK-CONDON EFFECT}

To reveal the physical essence of the FC effect in central spin system
with vertical transitions, we will present a schematic perspective
view for the collective-spin-based FC effect in comparison with the
conventional picture of the electron-photon FC phenomenon. 

As noted in the preceding section, when the system is at zero temperature,
the transition probability is modulated by the FC factor $f_{0\rightarrow n}$
(Eq.~\ref{transition_probability}). The most favored transition
must have the largest FC factor. And for a given hyperfine coupling
strength $A$, there is only one maximum or two adjacent maxima ($\left|f_{0\rightarrow n}\right|=\left|f_{0\rightarrow n+1}\right|$).
Hence, we require the most possible transition ($\left|\theta_{0},0;0\right\rangle $
to $\left|\theta_{1},n;1\right\rangle $) to satisfy
\begin{equation}
\begin{cases}
\left|f_{0\rightarrow n}\right|-\left|f_{0\rightarrow n+1}\right|\geqslant0,\\
\left|f_{0\rightarrow n}\right|-\left|f_{0\rightarrow n-1}\right|\geqslant0.
\end{cases}
\end{equation}
From Eq.~ (\ref{F-C_factor2}), we obtain the most favored final
Dicke state $\left|n_{\mathrm{mf}}\right\rangle $ by
\begin{equation}
n_{\mathrm{mf}}\leqslant\frac{N+1}{2}\left(1-\cos\theta\right)\leqslant n_{\mathrm{mf}}+1.\label{most_favored_n}
\end{equation}
Here $\theta$ is dependent on the coupling constant $A$ and $0\leqslant n_{\mathrm{mf}}\leqslant\left(N+1\right)/2$.
Thus the most favored transition is determined by hyperfine coupling
$A$ and number of the nuclear spins $N$.

\begin{figure}
\includegraphics[width=6cm]{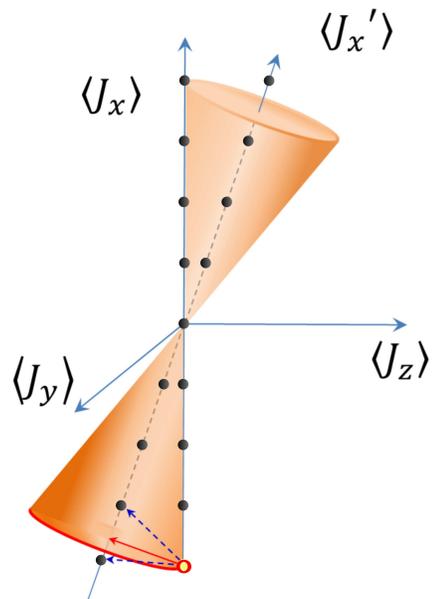}

\caption{\label{fig:vertical transition-2}(Color online) Time-evolution of
the nuclear spins, which was assumed to be initialized on the state
$\left|\theta_{0},0;0\right\rangle $, with small hyperfine coupling
$A=0.2\omega_{\mathrm{nu}}$.}
\end{figure}

In the space spanned by $\left\{ \left\langle J_{x}\right\rangle ,\ \left\langle J_{y}\right\rangle ,\ \left\langle J_{z}\right\rangle \right\} $,
the eigenstates of $H_{0}^{\left(0\right)}$ (Eq.~\ref{H0_eff})
are designated by the discrete dark gray dots on the $\left\langle J_{x}\right\rangle $
axis in Fig.~\ref{fig:vertical transition-2}. The initial state
of the collective nuclear spins ($\left|\theta_{0},0\right\rangle $)
is denoted by the bottommost red circle, with coordinate $\left\{ \left\langle J_{x}\right\rangle =-N/2,\ \left\langle J_{y}\right\rangle =0,\ \left\langle J_{z}\right\rangle =0\right\} $.
Let us consider the transition of the central spin from $\left|s=0\right\rangle $
to $\left|s=1\right\rangle $ induced by the polarization field. The
corresponding effective Hamiltonian of the nuclear spins after this
transition changes into
\begin{equation}
H_{0}^{\left(1\right)}=AJ_{z}+\omega_{\mathrm{nu}}J_{x}+D.
\end{equation}
After making a rotation around $y$-axis with angle $(-\theta)$,
we obtain
\begin{eqnarray}
H_{0}^{\prime\left(1\right)} & = & \tilde{\omega}_{\mathrm{nu}}J_{x}^{\prime}+D.
\end{eqnarray}
Its eigenstates distribute on the $\left\langle J_{x}^{\prime}\right\rangle $
axis as shown in Fig.~\ref{fig:vertical transition-2}. 

From Eq.~(\ref{most_favored_n}), the average of $J_{x}^{\prime}$
for the most favored state is given by 
\begin{equation}
\left\langle \theta_{1},n_{\mathrm{mf}};1\mid J_{x}^{\prime}\mid\theta_{1},n_{\mathrm{mf}};1\right\rangle \approx-\left(N+1\right)\cos\theta/2.\label{Jx}
\end{equation}
It is nearly the same as the projection of the initial value of $\left\langle J_{x}\right\rangle $
on the $J_{x}^{\prime}$ axis $-N\cos\theta/2$. In other words, during
the excitation of the central spin, the collective environmental spins
like to jump vertically from the initial state $\left|\theta_{0},0;0\right\rangle $
to the final state $\left|\theta_{1},n_{\mathrm{mf}};1\right\rangle $,
as depicted by the red arrow with a solid line in Fig.~\ref{fig:vertical transition-2}.
Usually, there is no such a eigenstates of the system corresponds
to this vertical transition. As a consequence, the two adjacent transitions
(characterized by the two blue arrows with dashed lines in Fig.~\ref{fig:vertical transition-2})
beside the vertical projection point are most likely to occur.

In the other hand, we investigate the dynamic evolution of the collective
spins. In the Heisenberg picture, the operators satisfy the Heisenberg
equations
\begin{equation}
\begin{cases}
\frac{\partial J_{x}^{\prime}}{\partial t}=0,\\
\frac{\partial J_{y}^{\prime}}{\partial t}=-\tilde{\omega}_{\mathrm{nu}}J_{z}^{\prime},\\
\frac{\partial J_{z}^{\prime}}{\partial t}=\tilde{\omega}_{\mathrm{nu}}J_{y}^{\prime}.
\end{cases}
\end{equation}
Since the transition of the central spin is very fast compared with
the motion of the nuclear spins. During the excitation of the electronic
spin from $\left|0\right\rangle $ to $\left|1\right\rangle $, the
average of the nuclear-spin operators can be assumed to be the same
before and after the transition, i.e., the geometry configuration
of the nuclear spins keep invariant. This is the essence of vertical
transition. Then we obtain corresponding initial mean values for the
new angular operators $\left\langle J_{x}^{\prime}\left(0\right)\right\rangle =-\left(N\cos\theta\right)/2$,
$\left\langle J_{y}^{\prime}\left(0\right)\right\rangle =0$, and
$\left\langle J_{z}^{\prime}\left(0\right)\right\rangle =-\left(N\sin\theta\right)/2$.
With these initial conditions, we obtain the solution of the former
differential equations as
\begin{equation}
\begin{cases}
\left\langle J_{x}^{\prime}\left(t\right)\right\rangle =-\frac{N}{2}\cos\theta,\\
\left\langle J_{y}^{\prime}\left(t\right)\right\rangle =\frac{N}{2}\sin\theta\sin\tilde{\omega}_{\mathrm{nu}}t,\\
\left\langle J_{z}^{\prime}\left(t\right)\right\rangle =-\frac{N}{2}\sin\theta\cos\tilde{\omega}_{\mathrm{nu}}t.
\end{cases}\label{Jx-1}
\end{equation}

As shown in Fig.~\ref{fig:vertical transition-2}, the trajectory
of the state of collective nuclear spins is a circle (the red one)
which crosses the initial point in the plane and is perpendicular
to the $\left\langle J_{x}^{\prime}\right\rangle $-axis. This intuitive
understanding of vertical transition lead to the same result as Eq.~(\ref{Jx}).
This confirms our former conjecture that the most favor final nuclear-spin
state after the transition is ruled by the vertical transition.

\section{Conclusion}

We studied the influence of the environmental spins on absorption
spectrum of the central spin. It is found that there exists similar
FC effect in the central spin model as that in conventional electron-phonon
model. In the zero temperature case, the original Lorentz absorption
spectrum of the bare central spin is shifted and split into few small
peaks. And the most probable transitions, which make the largest contribution
to the absorption spectrum, are governed by the ``vertical transition''
mechanism. If the system is at finite temperature, the peak of the
absorption spectrum is markedly depressed and broadened. Especially,
when the hyperfine coupling is strong enough, the excitation of the
central spin is intensively suppressed, which leads to the spin F-C
blockade.
\begin{acknowledgments}
We thank Da Zhi Xu and Cheng-Yun Cai for helpful discussion. This
work is supported by National Natural Science Foundation of China
under Grants No.11121403, No. 10935010 and No. 11074261.
\end{acknowledgments}
\appendix

\section{Rotated Dicke state}

We define the collective spin operators as
\begin{equation}
J_{\pm}=\sum_{j}I_{\pm}^{(j)},\ J_{\alpha}=\sum_{j}I_{\alpha}^{(j)},\ \mathrm{for}\ \alpha=x,y,z,
\end{equation}
where $I_{\pm}^{(j)}=I_{x}^{(j)}\pm iI_{y}^{(j)}$. It is easy to
find that the defined operators satisfy
\begin{equation}
[J_{i},J_{j}]=i\epsilon_{ijk}J_{k},\ [J^{2},J_{i}]=0,
\end{equation}
and
\begin{equation}
[J_{z},J_{\pm}]=\pm J_{\pm},\ [J_{+},J_{-}]=2J_{z},\ [J^{2},J_{\pm}]=0,
\end{equation}
where $\epsilon_{ijk}$ is the totally antisymmetric Levi-Civita tensor,
with $\epsilon_{xyz}=+1$. With the help of these operators, the collective
spins may be characterized by the simultaneous eigenstates $\left|J,M\right\rangle $
of $J^{2}$ and $J_{z}$ with \cite{Dicke state-1,Dicke states-2}
\begin{eqnarray*}
J & = & \frac{N}{2},\\
M & = & -\frac{N}{2},-\frac{N}{2}+1,\cdots,\frac{N}{2}-1,\frac{N}{2}.
\end{eqnarray*}
It is proofed that all of these states $\left|J,M\right\rangle $
(Dicke states) are symmetric under permutations of the nucleus and
all the symmetric states are in the subspace which is spanned by the
states with maximal angular momentum $J=N/2$. Now we order the totally
symmetric eigenstates as
\begin{eqnarray*}
\left|\frac{N}{2},\frac{N}{2}\right\rangle  & = & \left|1,1,\dots,1\right\rangle \\
\left|\frac{N}{2},\frac{N}{2}-1\right\rangle  & = & J_{-}\left|1,1,\dots,1\right\rangle =\mathcal{S}_{n}\left|0,1,\dots,1\right\rangle \\
\left|\frac{N}{2},\frac{N}{2}-2\right\rangle  & = & J_{-}^{2}\left|1,1,\dots,1\right\rangle =\mathcal{S}_{n}\left|0,0,1,\dots,1\right\rangle \\
\dots & \dots & \dots\\
\left|\frac{N}{2},-\frac{N}{2}\right\rangle  & = & J_{-}^{N}\left|1,1,\dots,1\right\rangle =\left|0,0,\dots,0\right\rangle ,
\end{eqnarray*}
where $\mathcal{S}_{n}$ is the symmetrization operator and $\left|N/2,-N/2\right\rangle $
the ground state of the atomic ensemble. 

Since all the operations are proceeding in the subspace of $J=N/2$,
we abbreviate the eigenfunction $\{\left|N/2,M=-N/2+m\right\rangle ,\ m=0,1,2,\dots,N\}$
of the operators $\{J^{2},J_{z}\}$ as 
\begin{equation}
\left|m\right\rangle \equiv\left|\frac{N}{2},-\frac{N}{2}+m\right\rangle ,
\end{equation}
which satisfy
\begin{equation}
\begin{cases}
J_{+}\left|m\right\rangle =\sqrt{\left(N-m\right)\left(m+1\right)}\left|m+1\right\rangle ,\\
J_{-}\left|m\right\rangle =\sqrt{m\left(N-m+1\right)}\left|m+1\right\rangle ,\\
J_{z}\left|m\right\rangle =\left(-\frac{N}{2}+m\right)\left|m\right\rangle .
\end{cases}
\end{equation}
It should be noted that we re-marked the eigenstate of the total angular
momentum, with the excitation number $m$ of the nuclear spins. 

The rotating operator corresponding to a rotation about the $y$ axis
reads as
\begin{equation}
R_{y}(\theta)=\exp[-i\theta J_{y}].\label{Ry}
\end{equation}
Since $J^{2}$ commutes with the rotation operators (i.e. $[J^{2},\ J_{y}]=0)$
and hence the subspace of $J=N/2$ is an invariant subspace of $R_{y}$.
After a unitary transformation via Eq.~(\ref{Ry}), we get the rotated
operators 
\begin{eqnarray}
J^{\prime2}=R_{y}\left(\theta\right)J^{2}R_{y}^{\dagger}\left(\theta\right) & = & J^{2},\\
J_{z}^{\prime}=R_{y}\left(\theta\right)J_{z}R_{y}^{\dagger}\left(\theta\right) & = & \cos\theta J_{z}+\sin\theta J_{x},
\end{eqnarray}
and the rotated eigenstate
\begin{eqnarray}
\left|\theta,m\right\rangle  & = & R_{y}\left(\theta\right)\left|m\right\rangle =\sum_{l}d_{l,m}^{N/2}\left|l\right\rangle ,
\end{eqnarray}
with the element of the Wigner's (small) d-matrix 
\begin{eqnarray}
 &  & d_{l,m}^{N/2}\left(\theta\right)=\left[m!(N-m)!l!\left(N-l\right)!\right]^{\frac{1}{2}}\nonumber \\
 &  & \times\sum_{k}\frac{\left(-\right)^{k}\left(\cos\frac{\theta}{2}\right)^{N+m-l-2k}\left(-\sin\frac{\theta}{2}\right)^{l-m+2k}}{\left(N-l-k\right)!\left(m-k\right)!\left(k+l-m\right)!k!}\!,
\end{eqnarray}
where $k=0,1,2,\dots$ and keep all the factorials non-negative.

\end{document}